\documentstyle[11pt,paspconf]{article}

\begin{document}

\title{ MEGA, a Wide-Field Survey of Microlensing in M31 }
\author{A.P.S.~Crotts, R.~Uglesich}
\affil{Department of Astronomy, Columbia University}
\author{G.~Gyuk}
\affil{Department of Physics, University of California, San Diego}
\author{A.B.~Tomaney}
\affil{Department of Astronomy, University of Washington}

\section{Introduction}

Microlensing has become an interesting probe of dark matter in our Galaxy.
Recent microlensing surveys have indicated that approximately one-half of the
matter in the Galactic halo may be dark objects with masses comparable to those
of stars, but have not revealed what these objects might be.
We have detailed how microlensing internal to M31 (Crotts 1992, Gyuk and Crotts
2000) might be used
to test such results and show better how the microlensing matter is
distributed in space and as a function of mass.
A survey of small fields in M31 has revealed several such candidate events, at
roughly the predicted rate (Crotts \& Tomaney 1996).
We discuss below what efforts have been required to further verify the
microlensing nature of these events.

Many papers have predicted the lensing optical depth $\tau$ in M31, which
should approach $\tau \approx 10^{-5}$, over an order of magnitude greater than
towards the LMC, but none of these works study the variation in halo lensing
optical depth over the face of M31.
Since M31 is distinct from the Galaxy in the many sightlines for microlensing
that it presents to an observer at Earth, the variation of $\tau$ depending on
the spatial distribution of lensing objects should be explored.

\section{Testing Candidate Microlensing Events}

Originally, Crotts \& Tomaney (1996) identified six events from the 1995
observing season in M31 (using the Vatican Advanced Technology 1.8-meter
telescope on Mt.~Graham, Arizona) over $\sim60$d covering a 120 arcmin$^2$
field.
These events were characterized by full-width half-maximum timescales of
10d$<t_{fwhm}<50$d.
The longer end of this range is troublesome, because it coincides with the
range of pulsewidths seen in miras and other longterm red variables.
In fact, a mira lightcurve, resembling a symmetric sawtooth in magnitude, with
peak-to-valley amplitudes of about $5^{mag}$ in the $R$ band, appears similar
to the lightcurve of a simple (point-mass, point-source) microlensing event
during its maximum amplification.
Furthermore, if the period of a mira is about $2 \over 3$~yr, one must monitor
for two M31 seasons past (or preceding) the peak in the light curve in order
to detect another peak, since proximate peaks occur when M31 is not easily
observable.
The peak of such a mira has $t_{fwhm}\approx 40$d, which would corresponds to
lensing masses $m \approx 0.5 M_\odot - 1 M_\odot$ for typical lensing
geometries in our survey region.
Hence mira-like variables are troublesome contributors to a potential false
event rate, and require multiple seasons of observations in order to be
eliminated.

We have performed two tests of these six original candidates: 1) constructing
well-sampled lightcurves over the M31 seasons of the three subsequent years
(through 1998),
and 2) obtaining $HST$ WFPC2 snapshot observations of these sources in order to
determine if their colors are consistent with miras-like variables.
(The latter test is impossible from the ground, since crowding does not allow
one to resolve typical sources in average seeing conditions, however,
variable sources are made to appear isolated from one another by virtue of
image subtraction e.g.~Tomaney \& Crotts [1996]).
The result of these two event filters is to eliminate three of the six
events, with the remainder firmly inconsistent with mira-like variables.
For the remaining events, now that we have measured their baseline magnitudes
from WFPC2 images, we can calculate a more accurate peak amplification
(assuming that the lensing mass rests as close as possible along the sightline
to the core of M31).
These persist in mass estimates for the lensing masses in the range $0.3M_\odot
\la m \la 1 M_\odot$, with two events possible arising from stars in M31's
bulge, but one almost certainly not a bulge lens, given its source position
2.5~kpc out into the disk.

\section{Possible Results from a Larger Survey}

Given the robust nature of at least some of the candidate microlensing events
seen in the small area survey discussed above, it is worthwhile considering the
possible outcome of a larger, wide-angle survey, especially considering the
advent of CCD imagers covering large fractions of a square degree.

We present here representative results simulating a survey in which roughly
one-half square degree is imaged for six hours every three nights on a
two-meter telescope, in 1-arcsec seeing, requiring each event to be sampled at
least twice at the 4$\sigma$ level.
The halo fraction, halo flattening ($q$) and core radius ($r_c$) are allowed to
vary between models, and then a maximum likelihood calculation is performed to
yield resulting values for these parameters.

Our calculations show that this larger survey might easily observe $\sim$100
such events per M31 observing season, which would allow the shape of the
microlensing halo of M31 to be mapped.
Since most masses reside near where the sightline passes the center of the
galaxy, at a known source-lens distance, this survey would also allow a more
exact determination of the masses doing the lensing.
Selected fields in M31 might also serve as an independent sightline through the
halo of our Galaxy.
The preliminary epochs for a large survey in M31, over one-half square degree,
have already been obtained, initiating the project MEGA: Microlensing
Exploration of the Galaxy and Andromeda.

All models produce $\ga$100 events per season.
Our ability to measure $r_c$ and $q$ depend on the true value of $r_c$, with
small values providing greater $\tau$ in the galaxy's center, where more
sources exist.
After three seasons, $r_c$ can be measured to within $\sim$1.5~kpc (1$\sigma$),
and $q$ to $\sim$0.1 (for $r_c < 5$~kpc), or $\sim$2.5~kpc and $\sim$0.2,
respectively for $r_c > 10$~kpc.
With $q$ and $r_c$ well-constrained, the data allow a
superior estimate of lens mass distribution.
This many events can result from a campaign using
existing wide-field CCD arrays on two-meter$+$ class telescopes.
We have initiated this effort (MEGA) by establishing long baselines
eliminating long-period variables, having
obtained several epochs of such data in 1997 and 1998, and having begun more
intensive observations in 1999 to detect microlensing events across much of
M31 over the course of several seasons.

\begin{figure}
\plotfiddle{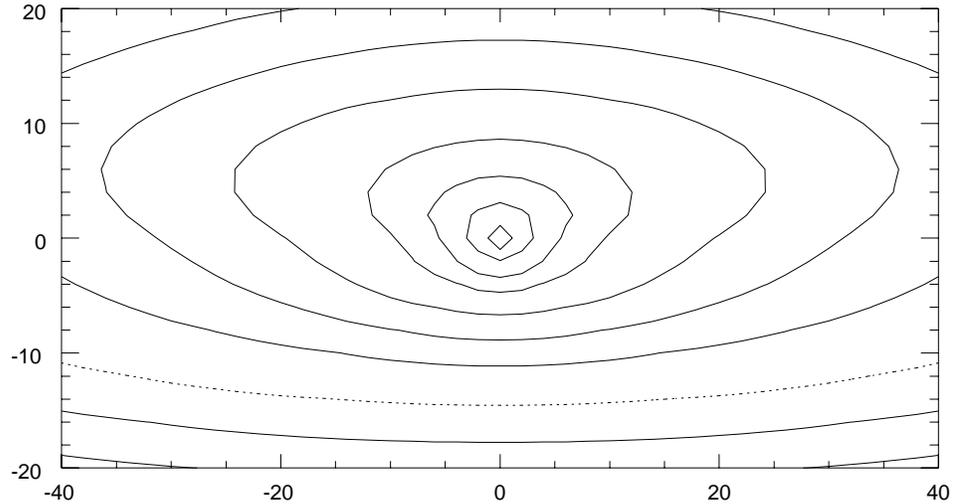}{3.3in}{-90}{060}{060}{-265}{330}
\caption{
An 80$\times$40 arcmin$^2$ plot of M31's center (major axis of M31 is
horizontal, minor axis vertical) showing contours of the predicted event rate
for bulge and halo microlensing in M31.
The highest contour is for 50 events yr$^{-1}$ arcmin$^{-1}$ (near the center),
with lower contours at 20, 10, 5, 2, 1, 0.5, 0.2 (dotted), 0.1 and 0.05 events
y$^{-1}$ arcmin$^{-1}$.
This reasonable model, for an unflattened halo with a 5~kpc core radius,
predicts over 100 detections during an M31 observing season.
}
\end{figure}

\begin{figure}
\plotfiddle{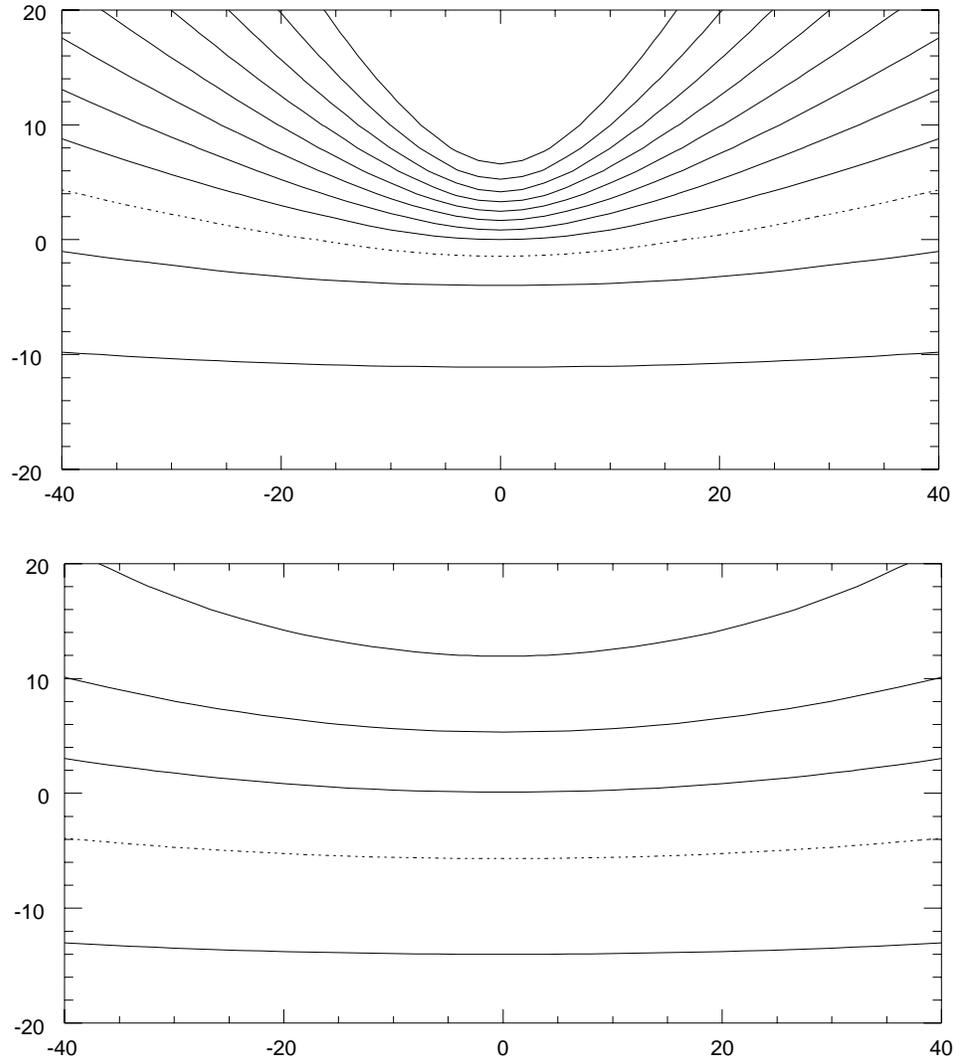}{5.5in}{-90}{060}{060}{-266}{540}
\caption{
The same region as in Fig.~1, but for contours of lensing optical
depth (halo only) for two models of the spatial distribution of halo
microlensing masses.
The top panel corresponds to an unflattened model with 1.5~kpc core
radius, and, on the bottom, a 3.3-to-1 flattening and a 10~kpc core radius.
The dotted contours (just below center in each panel) correspond to
$2\times 10^{-6}$, increasing towards the top (far side of
disk) to over $6\times 10^{-6}$ in the top panel and $3.5\times 10^{-6}$ in
the bottom.
In addition to the halo contribution, the central 10 arcmin diameter contains
a $\tau \la 5\times 10^{-6}$ bulge signal (represented in Fig.~1), a uniform
$\tau \la 10^{-6}$ due to the Galaxy, and a much smaller contribution from
M31 disk-disk lensing.
}
\end{figure}

\begin{figure}
\plotfiddle{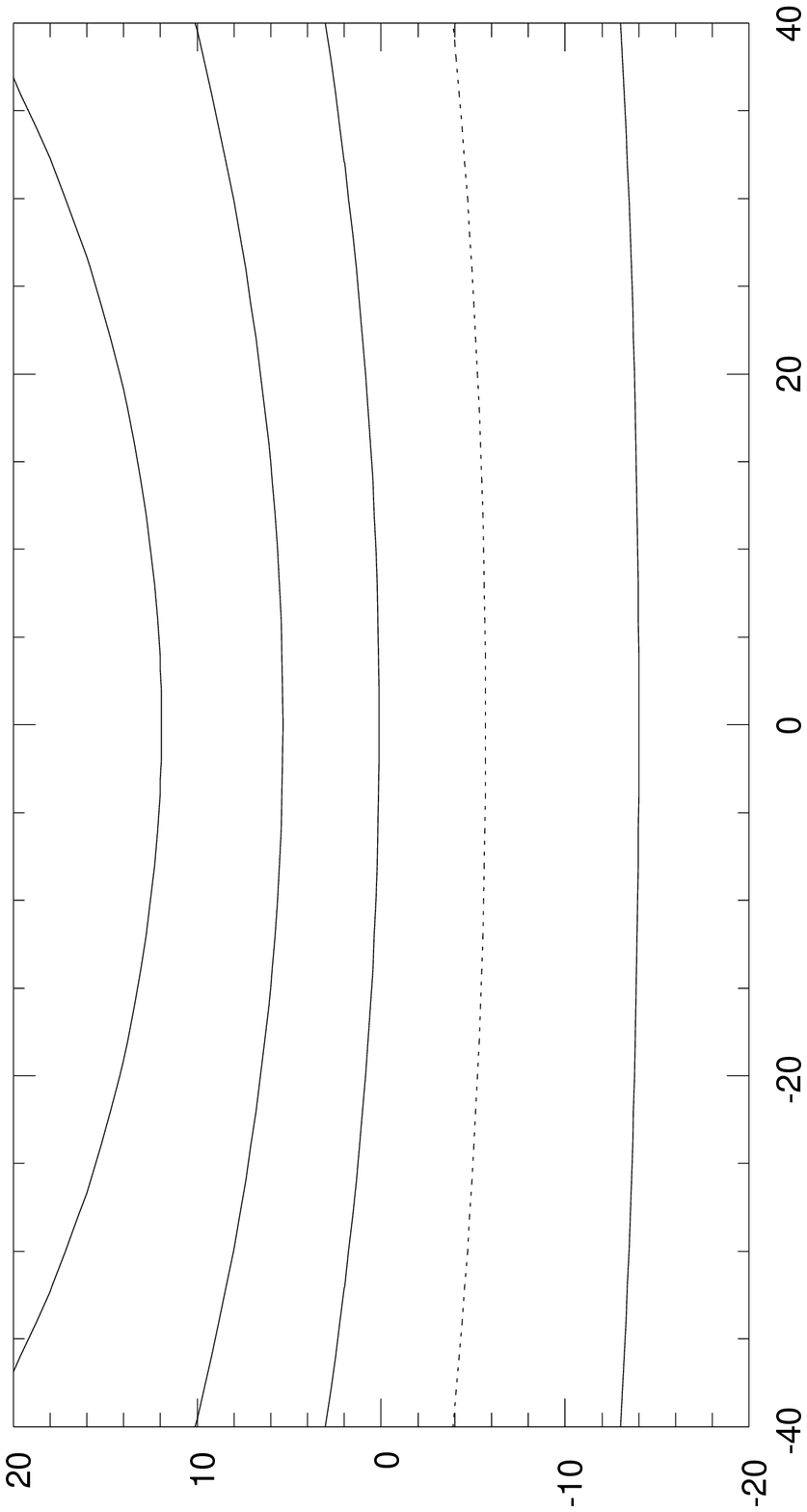}{0.0in}{-90}{060}{060}{-265}{510}
\end{figure}

\end{document}